\pdfoutput=1
\documentclass[aps,pra,10pt,twocolumn,showpacs,superscriptaddress,nofootinbib]{revtex4-1}

\usepackage[usenames,dvipsnames,svgnames]{pstricks}
\usepackage{epsfig}
\usepackage{pst-grad} 
\usepackage{pst-plot} 
\usepackage{epstopdf}
\usepackage{newlfont}
\usepackage{amssymb}
\usepackage{amsfonts}
\usepackage{amsmath}
\usepackage{amsthm}
\usepackage{graphicx}
\usepackage{epsfig}
\usepackage{color}
\usepackage{bm}
\usepackage{tikz}

\usepackage{times}

\begin{document}
\title {Relativistic quantum heat engine from uncertainty relation standpoint}
\author{Pritam Chattopadhyay and Goutam Paul}
\email{pritam.cphys@gmail.com, goutam.paul@isical.ac.in}
\affiliation{Cryptology and Security Research Unit, R.C. Bose Center for Cryptology and Security,\\
Indian Statistical Institute, Kolkata 700108, India}

\begin{abstract}
 Established heat engines in quantum regime can be modeled with various quantum systems as working substances. For example, in the non-relativistic case, we can model the heat engine using infinite potential well as a working substance to evaluate the efficiency and work done of the engine. Here, we propose quantum heat engine with a relativistic particle confined in the one-dimensional potential well as working substance. The cycle comprises of two isothermal processes and two potential well processes of equal width, which forms the quantum counterpart of the known isochoric process in classical nature. For a concrete interpretation about the relation between the quantum observables with the physically measurable parameters (like the efficiency and  work done), we develop a link between the thermodynamic variables and the uncertainty relation. We have used this model to explore the work extraction and the efficiency of the heat engine for a relativistic case from the standpoint of uncertainty relation, where the incompatible observables are the position and the momentum operators. We are able to determine the bounds (the upper and the lower bounds) of the efficiency of the heat engine through the thermal uncertainty relation. 
\end{abstract}
\maketitle

\section{Introduction}\label{section1}
In recent times, the study of thermal devices in the quantum regime~\cite{rk} have gathered more attraction for research. The various systems like the quantum amplifier~\cite{asta}, magnetic refrigerator and engines~\cite{muno}, semiconductor~\cite{acher}, thermoelectric generator~\cite{liu} and many other explore quantum laws. With the advent of quantum technology, the exploration of quantum heat engine has gathered more attraction like quantum Stirling cycle~\cite{yy,yy2}. They are conventionally used in the field of technology like power engineering and cryogenics.  Quantum heat engines have unfolded distinct techniques to extract work. The reason for this distinct path is quantum coherence and quantum entanglement which has no existence in the classical world. Numerous experiments to analyze quantum engine have been explored~\cite{jro,ht}.  Exploration of nano-scale devices like quantum ratchet~\cite{jja,rlo}, molecular motor~\cite{ber} and Brownian heat engine~\cite{iam,ldi} have enhanced the field of quantum thermodynamics.

The development of quantum information theory has made the study of quantum cycles interesting from an information standpoint. It was Maxwell who proposed a thought experiment~\cite{jcm} which expressed the correlation between information and thermodynamics. But his theory raised a contradiction with the existing thermodynamic laws. It was Leo Szilard, who proposed Szilard engine~\cite{lsz} and showed that the theory does not contradict the second law of thermodynamics. The extension of this engine in the quantum regime was proposed by Kim et al.~\cite{swk} which is dissimilar from the well known classical one.

Numerous quantum systems are considered for analysis of the quantum thermodynamics cycle, such as particles in a potential well~\cite{fwu,cou}, harmonic oscillator~\cite{lgc}, and spin 1/2 particles system~\cite{fwul}.  For example, quantum Szilard engine has been modeled using potential well.  One-dimensional infinite potential well~\cite{schiff,dj} is the simplest problem in non-relativistic quantum mechanics. This exemplifies how the wave nature of the particle quantizes the energy. When we place a barrier inside the middle of the well,  the single potential well gets converted to a two-chambered potential well i.e., a double potential well. The thorough analysis of this model has been shown in the work~\cite{thomas}. Now modeling relativistic heat engine using potential well is not so straightforward. As in relativistic quantum mechanics, the study of one dimensional potential well is not so straightforward. New features appear in the energy spectra due to spin and energy-momentum relation. The solution for the relativistic model of the potential well is shown while  keeping in mind that `Klein's paradox' is taken care of~\cite{ped}. Other problems that we face while we deal with the relativistic problem is the boundary conditions, which are not the same as in the case of non-relativistic problems. This is well discussed in~\cite{alo,menon}.

In this paper, we have first proposed a model which will exclusively work in the quantum regime for the relativistic scenario. So, for the analysis of the relativistic version of heat engine, we have considered one-dimensional potential well as the working substance.  In the next phase of our work, we establish a bridge between the uncertainty relation of position and momentum observable of the proposed model with our well-known thermodynamic variables.  So, the proposed model depicts an effective method for the analysis of the useful work without executing any measurement, but by using two reservoirs of different temperatures. The analysis of the work done by the engine has been explored from the uncertainty relation viewpoint where the incompatible observables are the position and the momentum operators of the relativistic particle in a potential well. 

With the advent of quantum information theory, the analysis and importance of uncertainty relation got enriched. It has numerous applications like, quantum cryptography~\cite{caf,koas,koas1}, entanglement detection~\cite{hofm,mart}, quantum metrology~\cite{giov} and quantum speed limit~\cite{marv,pires}. The thermal uncertainty relation that we have derived here is a special form of general uncertainty relation. The uncertainty relation of two observables is defined as
\begin{equation}
\Delta A \, \Delta B\geq \frac{\hbar}{2},
\end{equation}
where $A$, $B$ are two incompatible observable of the quantum regime. This relation states that no two incompatible observables can be measured with perfect accuracy in the quantum world. It can be measured with an accuracy which is of the order of Planck's constant ($\hbar$). So, uncertainty relation being a fundamental principle of quantum mechanics, the thermodynamic variables and models in the quantum regime needs to be bridged with this fundamental principle. Here, we have proposed a way to connect this fundamental principle with the thermodynamic engine models.

We have categorized the paper in this manner: in Section \ref{section2} we re-derive the uncertainty relation for a relativistic particle in a box using the Klein-Gordon equation.   Section \ref{section2A} is dedicated to the establishment of thermal uncertainty relation for a relativistic particle in the one-dimensional potential well of length $2L$. In Section \ref{section3}, we set-up the framework to develop the relationship between the thermodynamic variables with uncertainty relation for the relativistic particle in a potential well. Section \ref{section4} reveals the bound on the sum uncertainty relation for the relativistic particle from a thermal perspective.  Section \ref{section5} is devoted to discuss the Stirling cycle and then we develop the work done and efficiency from uncertainty viewpoint. Here, in this section, we generate the bound of work and efficiency of the relativistic quantum engine from the sum uncertainty relation. We concluded the paper in Section \ref{section6} with some discussion.

\section{Preliminaries}\label{section2}

\subsection{Revisiting uncertainty relation for relativistic particle in a potential well} 
Unlike the potential well problem in non-relativistic quantum mechanics, the potential well problem with a relativistic particle confined in it is not a textbook material traditionally. For our convenience we have used `$\equiv$' for defining a new symbol or quantity. 

Here, for the analysis, we have considered the relativistic potential well model as our working substance. The solution of the free Klein-Gordon (KG) equation~\cite{peskin} using Feshbach-Villars formalism~\cite{alb} is
\begin{eqnarray}\label{a}\nonumber
\psi_{\overrightarrow{p}}^{\pm} (\overrightarrow{x},t) & \equiv & A_{\pm} \begin{pmatrix}
\phi^{\pm} (\overrightarrow{p})\\[2mm]
\eta^{\pm}(\overrightarrow{p})
\end{pmatrix} e^{(\mp E_pt -\overrightarrow{p}\overrightarrow{x})/\hbar} \\ 
& = & A_{\pm} \phi^{\pm} (\overrightarrow{p}) e^{(\mp E_pt -\overrightarrow{p}\overrightarrow{x})/\hbar},
\end{eqnarray}
where $\pm$ represents the positive and negative energy solution respectively and $E_p = \sqrt{p^2c^2+m^2c^4}$, $A_{\pm}$ is the normalization constant and $m$, $p$, $c$ is the mass, momentum and the velocity (of the order of speed of light) of the particle respectively. 

The mathematical forms for $\phi^{\pm} (\overrightarrow{p})$ and $\eta^{\pm}(\overrightarrow{p})$  of Eq. ~\eqref{a} are given by
\begin{eqnarray}\label{b}\nonumber
\phi^{\pm} (\overrightarrow{p}) & \equiv & \frac{\pm E_p + mc^2}{2\sqrt{mc^2E_p}}\, , \\ \nonumber
\eta^{\pm}(\overrightarrow{p}) & \equiv & \frac{\mp E_p + mc^2}{2\sqrt{mc^2E_p}}. 
\end{eqnarray}

The procedure we generally take to solve for a particle in a box in the relativistic case leads to `Klein paradox'.  Klein paradox tells that the flux of the reflected plane wave in the walls of the potential well is larger than that of the incident waves. The reason behind this is the wavefunction which starts to pick up components from the negative energy states. The way to solve this paradox is to presume the mass of the system as a function of $x$. So it is defined as 
\begin{eqnarray}\label{c} \nonumber
m(x)\equiv \left\{
\begin{array}{lc}
  m, &\  x \in {L} ,\\
  M \to \infty, & \  x \notin {L},
\end{array}\right.
\end{eqnarray}
where $L$ is the length of the potential box.
So, the wave function inside the box results in 
\begin{equation}\label{d}
\Psi(x)\equiv  \sqrt{\frac{2}{L}}\phi^+(p) \sin(p\,x/\hbar).
\end{equation}
Here $pL=n\pi\hbar$ and $n=1,2,\cdots ,\infty$.
The quantized energy of the system takes the form 
\begin{eqnarray}\label{e}  \nonumber
E_n & \equiv & \sqrt{\frac{n^2\pi^2\hbar^2 c^2}{L^2} + m^2c^4 } \\
&\approx & mc^2 + \frac{n^2 \pi^2 \hbar^2}{2m L^2} + \cdots \,,
\end{eqnarray}
where in the last line a small $p/(mc) = n\pi \hbar/(Lmc)$ expansion is made. The second term arises by solving the Schr\"odinger equation. The $mc^2$ term represents the rest energy, and the dots represent second and higher order terms which are being neglected for our analysis.

Now, for our purpose we consider a relativistic particle of mass $m$ inside a one-dimensional potential box of length $2L$ with a bath at temperature $T$. We have considered the potential box of length $2L$ for calculation convenience when we insert a partition in the middle of the potential box. So, the wavefunction of the system for the $n$-th level, similar to the Eq. \eqref{d} which takes the form 
\begin{equation}\label{f}
\psi(x)= \sqrt{\frac{1}{L}}\phi^+(p) \sin(px/\hbar), \quad where \quad p\,(2L)=n\pi\hbar.
\end{equation}
So, the quantized energy of the considered system takes the form similar to Eq. \eqref{e} as 
\begin{eqnarray}\label{g}  \nonumber
E_n &=& \sqrt{\frac{n^2\pi^2\hbar^2 c^2}{(2L)^2} + m^2c^4 } \\ 
&\approx & mc^2 + \frac{n^2 \pi^2 \hbar^2}{2m (2L)^2} + \cdots \,. 
\end{eqnarray}
 
Having the information about the wavefunction and the eigenvalues, we are all set to analyze the uncertainty relation of the position and the momentum operator of the system.
The mathematical form of the uncertainty relation for the position and the momentum operator of the system is  
\begin{eqnarray}\label{h}   \nonumber
\Delta x \Delta p & \equiv & \sigma_x \sigma_p \\ \nonumber
& = & \frac{\hbar L}{2}\, \phi^+(p)\Big[\Big(\frac{1}{3}-\frac{2}{(n\pi)^2} - \phi^{+2}(p)\Big)  \\ \nonumber
& \times &  \Big(\frac{\pi^2 n^2}{L^2} +  \frac{8m^2c^2}{\hbar^2}\Big) \Big]^{\frac{1}{2}} \\ 
& \geq &  \frac{\hbar}{2}, 
\end{eqnarray}
where  $\Delta x^2 = \langle x^2 \rangle - \langle x \rangle^2$ and in the case of momentum, $\Delta p^2$ can be defined similarly. The mathematical form of the expectation values of  $\langle x \rangle$, $\langle p \rangle$, $\langle x^2 \rangle$ and $\langle p^2 \rangle$ for the relativistic particle confined in the potential well  are 
\begin{eqnarray} \label{i} \nonumber
\langle \psi_n|p|\psi_n \rangle & = & 0 \\\nonumber
\langle \psi_n|x|\psi_n\rangle & = & L\, \phi^{+2}(p) \quad    n=1,2,\cdots \\ \nonumber
\langle \psi_n|x^2|\psi_n\rangle & = & 4L^2\,\phi^{+2}(p)\Big[\frac{1}{3}- \frac{1}{2(n\pi)^2}\Big]\,\, n=1,2,\cdots \\ 
\langle \psi_n|p^2|\psi_n\rangle & = & \Big(\frac{\pi \hbar n}{2L}\Big)^2 + 2m^2 c^2,  \quad     n=1,2,\cdots  
\end{eqnarray}
where we have considered the wavefunction $\psi_n$ as shown is Eq. \eqref{f}.

\section{Results} 

\subsection{Thermal uncertainty relation for relativistic particle}\label{section2A}
Now, we will formulate the uncertainty relation of this system from the thermodynamic standpoint. To evaluate the thermal uncertainty relation we have to analyze the partition function of the system. The partition function~\cite{reif1}, $Z$, for 1-D potential well where a relativistic particle is confined in it is expressed as
\begin{equation} \label{zzz}
Z  \equiv  \sum_{n=1}^{\infty} e^{-\beta E_n}  
 \approx  \frac{1}{2} \sqrt{\frac{\pi}{\beta \alpha}}\, e^{-\beta m c^2}\,, 
\end{equation}

where $\beta =\frac{1}{k_BT}$, $k_B$ being Boltzmann's constant and $\alpha = \frac{\pi^2 \hbar^2}{2m (2L)^2}$. The expression of $Z$ takes this form as the product of $\beta$ and $\alpha$ is a small quantity. The expectation of the $n$-th state of the system is 
\begin{equation}\label{nn}
\bar{n}   \equiv  \frac{\sum_n n e^{- \beta E_n}}{\sum_n e^{-\beta E_n}}   \approx \frac{1}{\sqrt{\pi \alpha \beta}}\,.
\end{equation}

After the evaluation of the partition function of the system,  we now have all the available resources to develop the thermal uncertainty relation for the relativistic particle in a 1-D potential well. 
So, to evaluate the uncertainty relation for the position and the momentum operator we have to calculate the variance of the position and the momentum operator for this system. For the evaluation of the expectation of the position operator we consider the $n$-th state of the system and using Eq.~\eqref{i} we get 
\begin{eqnarray} \label{l} \nonumber
 (\Delta X)^2_T & \equiv & \langle (\Delta X)^2 \rangle_T = \langle X^2\rangle_T - \langle X \rangle^2_T \\ \nonumber
  & \equiv & \frac{1}{Z} \Big( \sum_{n=1}^{\infty} \langle \psi_n|X^2|\psi_n\rangle e^{-\beta E_n} - \sum_{n=1}^{\infty} \langle \psi_n|X|\psi_n\rangle e^{-\beta E_n} \Big)  \\ \nonumber 
 & = & - \frac{2L^2}{\pi^2} \phi^{+2}(p) \, \frac{e^{-\alpha\beta} - \sqrt{\pi \alpha \beta} \times  erfc(\sqrt{\alpha\beta})}{\frac{1}{2} \sqrt{\frac{\pi}{\alpha \beta}}} \\ \nonumber
 & + & \frac{4L^2}{3} \phi^{+2}(p) -L^2 \phi^{+4}(p)  \\ \nonumber
 & = & - \phi^{+2}(p) \, \frac{4L^2 \sqrt{\alpha \beta}}{\pi^{5/2}} \times (e^{-\alpha \beta}- \sqrt{\pi \alpha\beta}) \\ 
 & + & L^2 \, \phi^{+2}(p) \Big(\frac{4}{3}-\phi^{+2}(p)\Big).
\end{eqnarray}
Here, $erfc$ is the complementary error function~\cite{andrew},  which emerges while evaluating the expression  $\langle X^2\rangle$. 

Similar to the expression of the dispersion relation of the position operator, the variance of the momentum  operator is 
\begin{eqnarray} \label{m} \nonumber
 (\Delta P)^2_T & \equiv & \langle (\Delta P)^2 \rangle_T = \langle P^2\rangle_T - \langle P \rangle^2_T \\  \nonumber
 & \equiv & \frac{1}{Z} \sum_{n=1}^{\infty} \langle \psi_n|P^2|\psi_n\rangle e^{-\beta E_n} \\ 
 & = & \frac{\pi^3 \hbar^2 \bar{n}^2}{8 L^2} + 2mc^2.  
\end{eqnarray}
So the uncertainty relation from Eq.~\eqref{l} and Eq.~\eqref{m}, at a thermal condition for the potential well model is expressed as
\begin{eqnarray} \label{n}\nonumber
 \Delta X_T \, \Delta P_T 
 & = & \frac{\hbar}{2} \Bigg[- \phi^{+2}(p) \, \frac{4L^2 \sqrt{\alpha \beta}}{\pi^{5/2}} \times (e^{-\alpha \beta}- \sqrt{\pi \alpha\beta})  \\ \nonumber
&  + & L^2 \, \phi^{+2}(p) \Big(\frac{4}{3}-\phi^{+2}(p)\Big) \Bigg]^{1/2} \\ \nonumber
 & \times & \Big(\frac{8mc^2}{\hbar^2} + \frac{\pi^3 \bar{n}^2}{2L^2} \Big)^{\frac{1}{2}} \\
 & \geq & \frac{\hbar}{2}\,. 
 \end{eqnarray}
 Along with the product uncertainty relation, we also evaluate the thermal sum uncertainty relation of the position and the momentum operator for the potential well problem. Here, we have calculated the sum uncertainty as we are concerned about the fact that the product uncertainty relation is unable to capture the uncertainty of the incompatible observables when the wavefunction is an eigenfunction of one of the observables.
The sum of uncertainty for these observables is 
 \begin{eqnarray} \label{0}\nonumber
 \Delta X_T + \Delta P_T 
 & = & \Bigg[- \phi^{+2}(p) \, \frac{4L^2 \sqrt{\alpha \beta}}{\pi^{5/2}} \times (e^{-\alpha \beta}- \sqrt{\pi \alpha\beta})  \\ \nonumber
 & + & L^2 \, \phi^{+2}(p) \Big(\frac{4}{3}-\phi^{+2}(p)\Big) \Bigg]^{1/2}  \\ \nonumber
 & + & \frac{\hbar}{2} \Big(\frac{8mc^2}{\hbar^2} + \frac{\pi^3 \bar{n}^2}{2L^2} \Big)^{1/2}\\
 & \geq & \frac{\hbar}{2}\, .
 \end{eqnarray}
 \begin{figure}[h]
 \begin{center}
 \includegraphics[width=1.0\columnwidth]{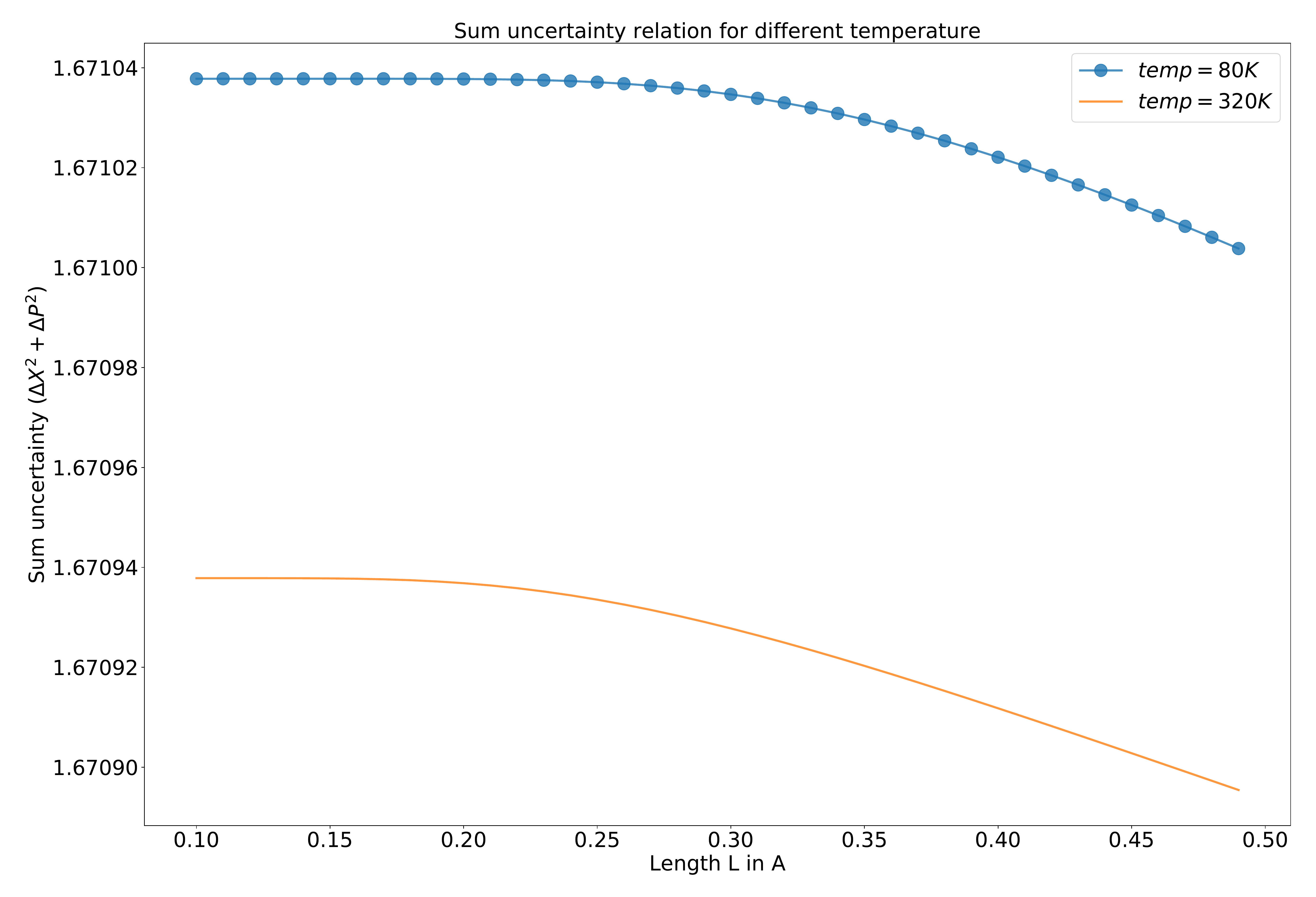}
 \end{center}
  \caption{The variation of sum uncertainty relation for different temperature. The dotted line is for lower and the solid line is for higher temperature.}
  \label{fig1}
  \end{figure}
   Fig.\ref{fig1} describes the variation of uncertainty with respect to different temperatures. The value of the uncertainty relation is almost constant for lower values of the length of the well. There is a sudden drop in the measure of uncertainty of the observables as the length of the well exceeds from $0.3$ $ \AA$. The dip is more for higher values of $L$.
  
  \begin{figure}[h]
  \begin{center}
   \includegraphics[width=1.0\columnwidth]{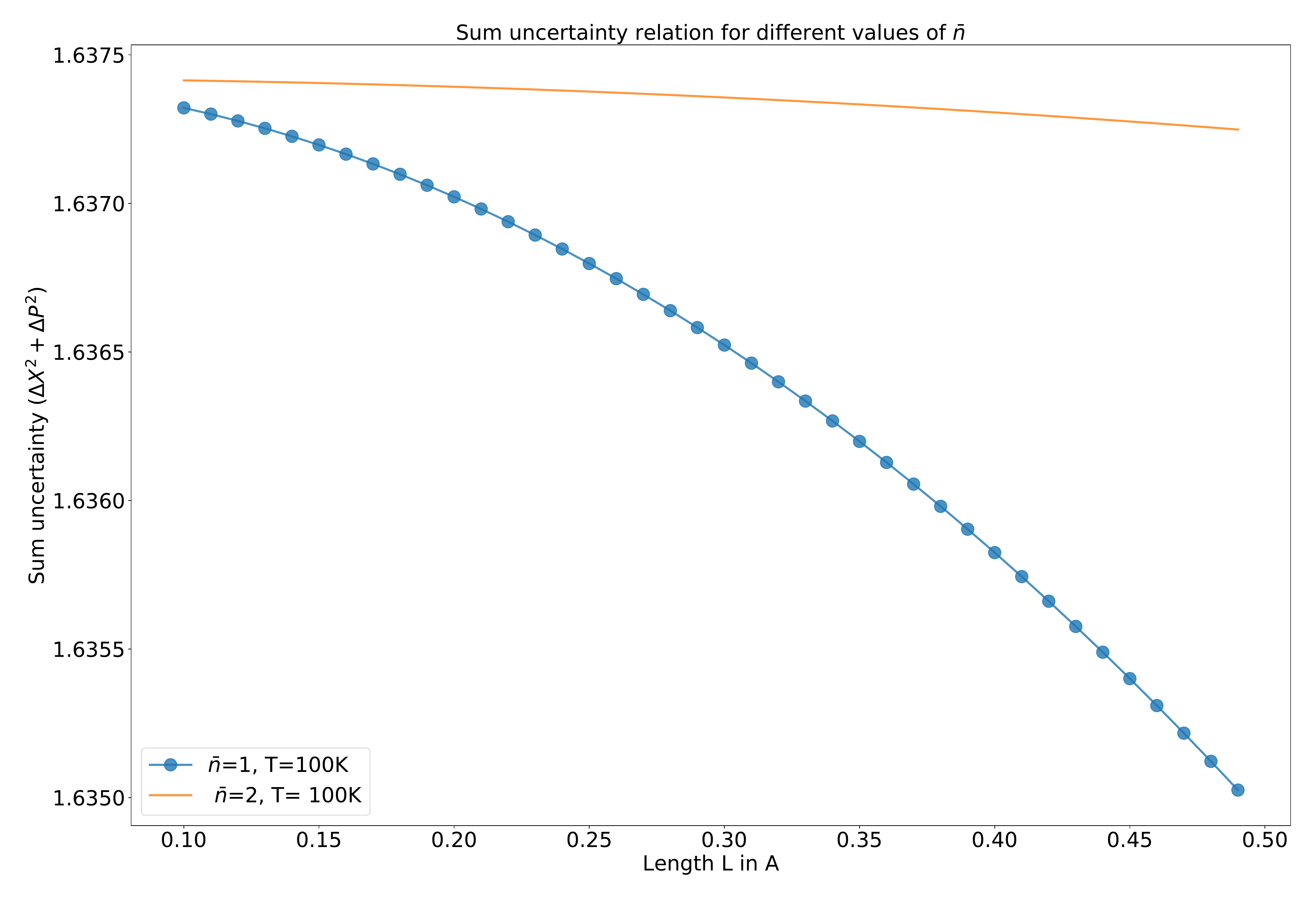}
  \end{center}
  \caption{This shows the variation of sum uncertainty relation for different values of $n$.}
  \label{fig2}
  \end{figure}
In Fig. \ref{fig2}, we can see that there is almost a gradual fall in the measure of uncertainty for $\bar{n}=1$. Whereas, for $\bar{n}=2$, we can visualize a small change for higher values of $L$. 

For the analysis of Fig.~\ref{fig1}, we have replaced $\bar{n}$ from Eq. \ref{nn} in the expression of  Eq. \ref{0}. Whereas, for the analysis of Fig.~\ref{fig2}, we have replaced the required term of Eq. \ref{0} as a function of $\bar{n}$ using Eq. \ref{nn} (for a fixed temperature `$T=100K$'), to have a clear understanding of the dependency of the uncertainty relation with temperature and the average `$n$'.
 
In our model, the particle is confined to a box of length `$2L$'. The uncertainty in the position is a function of `$L$', i.e., the particle has to be somewhere in the box. So, with the increase in length, there is an increase in the ``uncertainty of position", i.e., $\Delta X$ increases with increase in length. From the definition of Heisenberg uncertainty, the uncertainty of momentum decreases with length, as it is inversely proportional to the length. So, according to Heisenberg's definition, the overall uncertainty remains a constant (i.e $\hbar/2$). Following the same ideology in case of our analysis, the ``uncertainty of position" ($\Delta X$ of Eq. \ref{l}) should show more dominance over the contribution of $\Delta P$ for an increase in the length of the potential well in Eq. \ref{0}. We encounter a decrease in the ``total uncertainty" for higher values of $L$ which is depicted in Fig. \ref{fig1}. The reason behind this is the dominance of the first term of the expression of $\Delta X$ over the second term in Eq. \ref{l} due to its exponential nature, which causes an overall decrease in the ``total uncertainty". We encounter the same nature in Fig. \ref{fig2}. The reason for this nature is obviously similar to the analysis made for Fig. \ref{fig1}.

\subsection{Correlation of thermodynamic quantities with uncertainty of relativistic particle} \label{section3}
As far as our knowledge, the expression of the thermodynamic quantities from uncertainty relation for a relativistic particle has not yet been provided. We have developed the relationship between the basic thermodynamic quantities with the variance of the position and the momentum operator.

The partition function~\cite{reif1} of the system, $Z$, in terms of the variance by using Eq. ~\eqref{0} for replacing $\bar{n}$ in Eq.~\eqref{zzz} is expressed as
\begin{eqnarray}\label{eq1} 
Z =  \frac{\pi}{2} e^{-\beta mc^2} \, \Big[\frac{16c\sqrt{2mc}}{\pi^3 \hbar^2}(\Delta X_T + \Delta P_T + C_T)\Big]^{\frac{1}{2}},
\end{eqnarray}
where $C_T = L \phi \sqrt{(\frac{4}{3}-\phi^2)} \Big[\frac{2(\alpha \beta -\sqrt{\pi}(\alpha \beta)^{3/2}-1)}{\pi^{5/2}\sqrt{\alpha \beta}(\frac{4}{3}-\phi^2)}-1 \Big]-\sqrt{2mc}$. Similarly, the internal energy of the canonical system can be evaluated using the variance of two incompatible operators. For our analysis these two incompatible operators are the position and the momentum operator. The internal energy of the system from Eq. ~\eqref{eq1} evolves to 
\begin{eqnarray}\label{eq2} \nonumber
\langle E \rangle & \equiv & - \partial ln Z \big/ \partial \beta \\ \nonumber
& = & mc^2 + \frac{\zeta + \eta}{\pi [\frac{16c\sqrt{2mc}}{\pi^3 \hbar^2}(\Delta X_T + \Delta P_T + C_T)\Big]}\, ,
\end{eqnarray} 

where $\zeta$ is expressed as $\zeta= \frac{16c\sqrt{2mc}}{\pi^3 \hbar^2} \Big[\frac{2L\phi}{\pi^{5/2}\sqrt{\alpha \beta (\frac{4}{3}-\phi^2)}}(\alpha-\alpha^{3/2}\sqrt{\pi \beta})- \frac{L\phi}{\pi^{5/2}\beta^{3/2} \sqrt{\alpha(\frac{4}{3}-\phi^2)}} (\alpha \beta -\sqrt{\pi}(\alpha \beta)^{3/2}-1) \Big]$ and after calculation $\eta$ is conveyed as $\eta = \frac{\frac{4L^2\phi^2}{\pi^{5/2}}\Big[\sqrt{\frac{\alpha}{4\beta}} \Big(e^{-\alpha \beta}-\sqrt{\pi \alpha \beta}\Big) + \sqrt{\alpha \beta} \Big(-\alpha e^{- \alpha \beta}-\sqrt{\frac{\pi \alpha}{4\beta}}\Big) \Big]}{2 \Big[ L^2\phi^2 \Big(\frac{4}{3}-\phi^2 \Big)- \frac{4L^2\phi^2\sqrt{\alpha \beta}}{\pi^{5/2}} \Big(e^{-\alpha \beta} - \sqrt{\pi \alpha \beta}\Big)\Big]^{\frac{1}{2}}}$.

Having the information of the link between the uncertainty relation and the partition function of the system we are set to describe all the thermodynamic variables in terms of the uncertainty relation of the position and the momentum operator of the considered system.
One of the basic thermodynamic quantity is Helmholtz free energy~\cite{reif1} `F'. The Helmholtz free energy for the relativistic particle in terms of the uncertainty relation is  
\begin{eqnarray} \label{eq3} \nonumber
F & \equiv & \frac{-1}{\beta} ln Z \\ \nonumber
& = & mc^2 - \frac{1}{\beta}\, ln \Big[\frac{4c\sqrt{2mc}}{\pi \hbar^2}(\Delta X_T + \Delta P_T + C_T)\Big]^{\frac{1}{2}}. 
\end{eqnarray}

We know that we can define entropy from Helmholtz free energy. So, we are now able to express entropy in terms of uncertainty relation which is expressed as
\begin{eqnarray} \label{eq4} \nonumber
S & \equiv & - \frac{\partial F}{\partial T} \\ \nonumber 
& = & k_B\,\,ln\Big[\frac{4c\sqrt{2mc}}{\pi \hbar^2}(\Delta X_T + \Delta P_T + C_T)\Big]^{\frac{1}{2}} \\ 
& + &  \frac{\tau + \chi}{\pi \beta [\frac{16c\sqrt{2mc}}{\pi^3 \hbar^2}(\Delta X_T + \Delta P_T + C_T)\Big]}\, ,
\end{eqnarray}
where $\tau$ is expressed as $\tau= \frac{16c\sqrt{2mc}}{\pi^3 \hbar^2} \Big[\frac{2L k_B\phi}{\pi^{5/2}\sqrt{\alpha \beta (\frac{4}{3}-\phi^2)}}(\alpha-\alpha^{3/2}\sqrt{\pi \beta})- \frac{L k_B \phi}{\pi^{5/2}\beta^{3/2} \sqrt{\alpha(\frac{4}{3}-\phi^2)}} (\alpha \beta -\sqrt{\pi}(\alpha \beta)^{3/2}-1) \Big]$ and the form of $\chi$ after evaluation (using Eq.~\eqref{0} and the definition of $C_T$ defined in Eq.~\eqref{eq1}) is\\ $\chi = \frac{\frac{4L^2 k_B \phi^2}{\pi^{5/2}}\Big[\sqrt{\frac{\alpha}{4\beta}} \Big(e^{-\alpha \beta}-\sqrt{\pi \alpha \beta}\Big) + \sqrt{\alpha \beta} \Big(-\alpha e^{- \alpha \beta}-\sqrt{\frac{\pi \alpha}{4\beta}}\Big) \Big]}{2 \Big[ L^2\phi^2 \Big(\frac{4}{3}-\phi^2 \Big)- \frac{4L^2\phi^2\sqrt{\alpha \beta}}{\pi^{5/2}} \Big(e^{-\alpha \beta} - \sqrt{\pi \alpha \beta}\Big)\Big]^{\frac{1}{2}}}$.

\begin{figure}[h]
\begin{center}
\includegraphics[width=1.0\columnwidth]{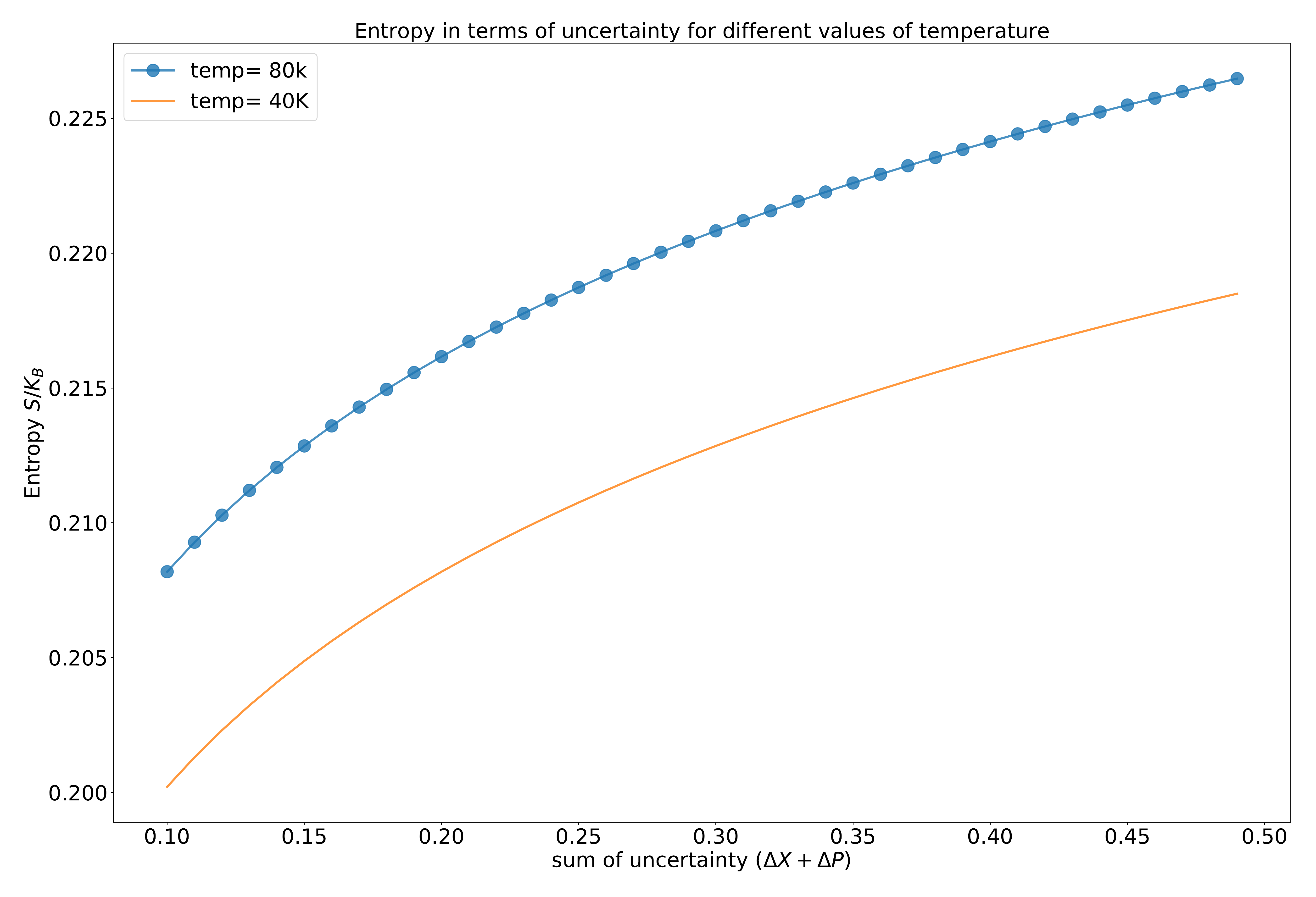}
\end{center}
  \caption{The variation of entropy from Eq.~\eqref{eq4} for different temperature is shown. The scattered plot is for higher temperature and the solid line is for lower temperature.}
  \label{fig3}
  \end{figure}
From Fig. \ref{fig3}, we can infer that the entropy of the system increases along with the increase of the uncertainty of the observables. This is true when the system is kept at a different temperature.

Till now entropy is the best-known measuring tool for entanglement. There is so far no standard method for the measure of entanglement for mixed states. If we can bridge a connection between these two quantities then it raises a question whether this can be a standard method for the entanglement measure.

For a given thermodynamic system, the knowledge of the Helmholtz free energy $F$ is enough for determining all other thermodynamic variables for the given system. Here, we have developed the correlation of Helmholtz free energy with the uncertainty relation of the position and the momentum operator of the relativistic particle.  This helps us to overcome the explicit requirement of internal energy of the system for the analysis of quantum thermodynamic system from uncertainty viewpoint. We can also explore and develop a theory which can explain the phase transition for relativistic particles in terms of their uncertainty relation. This is an open area to explore in the near future.

\subsection{Bound on sum uncertainty for relativistic model of one dimensional potential well}\label{section4}
The thorough analysis of the product uncertainty which produce better lower and upper bound using the method proposed in previous works~\cite{pc,mondal} results to zero. So, the product of variances of the specified observables is unable to capture the uncertainty for two incompatible observables. The reason behind this result is that the state of the system is an eigenstate of one of the observables. This causes the product of the uncertainties to vanish. We can overcome this issue if we invoke the sum of variances to capture the uncertainty of two incompatible observables. For the relativistic 1-D potential well, the sum of variance of two incompatible observable which results to the lower bound is defined as
\begin{equation} \label{eqn1}
\Delta A^2 +\Delta B^2 \geq \frac{1}{2} \sum_n \Big(\Big|\langle\psi_n|\bar{A}|\psi\rangle\Big|+\Big|\langle\psi_n|\bar{B}|\psi\rangle\Big|\Big)^2.
\end{equation}
 Here we replace $A$ by $X$ and $B$ by $P$, according to the system we have considered for our analysis. This results to the upper bound of the relation for position and momentum. It is expressed as
\begin{equation} \label{eqn2}
\Delta X^2+ \Delta P^2 \geq \frac{1}{2} \sum_n \Big(\Big|\langle\psi_n|\bar{X}|\psi\rangle\Big|+\Big|\langle\psi_n|\bar{P}|\psi\rangle\Big|\Big)^2.
\end{equation}

We can develop the upper bound of uncertainty relation for two incompatible observables when we compute the reverse uncertainty relation. We utilize the Dunkl-Williams inequality~\cite{pce} to evolve the reverse relation.
 The mathematical form of the inequality is 
\begin{equation} \label{eqn3}
\Delta A + \Delta B\leq \frac{\sqrt{2}\Delta (A-B)}{\sqrt{1-\frac{Cov(A,B)}{\Delta A. \Delta B}}}.
\end{equation}
 Squaring both sides of the Eq.~\eqref{eqn3} we get 
\begin{equation}\label{eqn4}
\Delta A^2 + \Delta B^2 \leq \frac{2 \Delta(A-B)^2}{1-\frac{\textrm{Cov}(A,B)}{\Delta A \Delta B}} - 2 \Delta A \Delta B\, ,
\end{equation}
where $Cov(A,B)$ is defined as $Cov(A,B) \equiv \frac{1}{2}\langle\{A,B\}\rangle-\langle A \rangle \langle B \rangle,$ and $\Delta (A-B)^2 \equiv \langle (A-B)^2 \rangle - \langle (A-B) \rangle^2.$

Now, for the  system which we have considered as our working substance, we calculate the reverse relation for the position and the momentum operator. So, we substitute $A$ by $X$ and $B$ by $P$ in  Eq.~\eqref{eqn4} which stands as
\begin{eqnarray}\label{eqn7} \nonumber
\Delta X^2 +\Delta P^2 & \leq & \frac{2 \Delta(X-P)^2}{1-\frac{Cov(X,P)}{\Delta X \Delta P}}- 2 \Delta X \Delta P \\ \nonumber
& \leq & 4L^2 \phi^{+2}(p) \,\Bigg( \frac{1}{3} - \frac{1}{2(n\pi)^2}\Bigg) \\ 
 & +  & \frac{\pi^2\hbar^2 n^2}{4 L^2} + 2m^2c^2\, . 
\end{eqnarray}
In Eq. \eqref{eqn7}, we have illustrated the reverse relation of the sum uncertainty relation without taking the thermal state under consideration.
Now, we evaluate the reverse sum uncertainty relation from the correlation of the thermal variables. The mathematical form for the relation stands as
\begin{eqnarray}\label{eqn8}  \nonumber
\Delta X_T^2 +\Delta P_T^2 
& \leq &  - \frac{8L^2 \sqrt{\alpha \beta}}{\pi^{5/2}} \,\phi^{+2}(p) (e^{-\alpha \beta}- \sqrt{\pi \alpha\beta}) \\ \nonumber
& + & \frac{8L^2}{3}\, \phi^{+2}(p) - 2L^2\phi^{+4}(p) \\ 
&  + & \frac{\hbar^2\bar{n}^2 \pi^3}{4L^2}\, + 4mc^2. 
\end{eqnarray} 

The Eq. \eqref{eqn8} express the upper bound of the sum uncertainty relation for our potential well model from the thermodynamic standpoint.

\subsection{Relativistic Stirling cycle and bound on it's efficiency}\label{section5}
\begin{figure}
\center
\includegraphics[width=1.0\columnwidth]{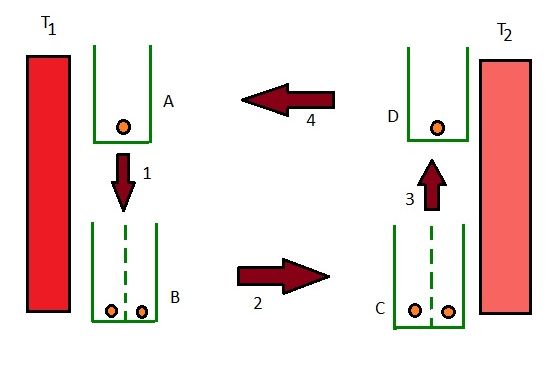}
\caption{The figure constitutes of four stages of the Stirling cycle of relativistic particle which is  modeled by one dimensional potential well.}
 \label{fig'}
\end{figure}

Here we consider the Stirling cycle for a relativistic particle. A Stirling cycle~\cite{say, agar,hua, blick}, comprises of four stages, where two are isothermal processes and the other two are isochoric processes.
In the first stage of the cycle, we place a barrier in the middle of the potential well isothermally, having a relativistic particle in it. The insertion of the barrier in the middle of the infinite potential well, which is represented by a delta potential, converts the single potential well to an infinite double potential well. Here, for our analysis, we consider a delta potential growing in strength from zero to a height which is large enough to prevent any tunneling through the barrier. So, it ensures us that the probability of tunneling through the barrier tends to zero if the tunneling time is more than the time required to complete the thermodynamic processes. The working medium which is connected with a hot bath (temperature $T_1$) remains at equilibrium condition during the quasi-static insertion. An isochoric heat extraction is experienced by the working medium when connected to a bath at a temperature $T_2$ where $T_2< T_1$. During the next stage of the cycle, the barrier is removed isothermally. While this process is carried out, the engine remains in equilibrium at temperature  $T_2$. We observe isochoric heat absorption in the final stage of the cycle when the system is reunited to the bath at temperature $T_1$. The pictorial representation of the cycle is shown in Fig.~\ref{fig'}.

In the work~\cite{thomas}, they have analyzed work done and efficiency for the heat engine in the non-relativistic limit. Here we have first developed heat engine in the relativistic limit where the working substance is the one dimensional potential well. Following the similar methodology, we have analyzed the work done and the efficiency for the heat engine for a  relativistic particle. Along with that, we develop the work done by the engine and its efficiency from the uncertainty relation viewpoint. We have considered a one dimensional potential well of length $2L$  with a relativistic particle of mass $m$ at temperature $T_1$ as the working substance for our analysis. The energy for the system is equivalent to Eq. ~\eqref{g}. The partition function of our system is $Z_A  =  \sum_{n=1}^{\infty} e^{-\beta E_n} \, \approx \Big(\frac{1}{2} \sqrt{\frac{\pi}{\beta \alpha}}\, e^{-\beta m c^2}\Big)$.
 Now, when we insert a wall isothermally it converts the one-dimensional infinite potential well into an infinite double well potential. In this situation, the energy level for even values of $n$ remain unchanged but we see a shift for the odd ones. It overlaps with their nearest neighbor energy level.
 The energy of the one-dimensional potential box that are created due to the partition is
\begin{equation}\label{eqnn1}
E_{2n} = \frac{(2n)^2\pi^2 \hslash^2}{2m(2L)^2} + mc^2,
\end{equation}
which is evaluated by replacing $n$ by $2n$ in Eq.~\eqref{g}.
The partition function for the newly formed partitioned potential well equivalent to Eq.~\eqref{zzz} is
\begin{eqnarray}\label{eqnn2} \nonumber
Z_B = \sum_n 2e^{-\beta_1 E_{2n}}. 
\end{eqnarray}

 The internal energy $U_A$ and $U_B$ is defined as $U_i \equiv - \partial ln Z_i \big/ \partial \beta_1$, where $i=A,B$ and $\beta_1= \frac{1}{k_BT_1}$. So, the internal energy are
\begin{equation}\label{eqnn3} 
U_A= U_B= \frac{1}{2\beta_1} + mc^2.
\end{equation}
 During the isothermal process, the heat exchange is expressed as 
\begin{equation}\label{eqnn4}
Q_{AB} \equiv U_B - U_A + k_B T_1 ln Z_B - k_B T_1 ln Z_A.
\end{equation}
After the isothermal process, the system is connected to a heat bath at temperature $T_2$. The partition function for this stage of the cycle is
\begin{equation}\label{eqnn5} \nonumber
Z_C = \sum_n 2e^{-\beta_2 E_{2n}}.
\end{equation}
In the second stage of the cycle, the heat exchanged  is given by the difference  of  the  average  energies  of  the  initial  and  the  final states (similar to Eq.~\eqref{eqnn3}). It is expressed as
\begin{equation}\label{eqnn6} 
Q_{CB} = U_C- U_B.
\end{equation}
Here $U_C= - \partial ln Z_C\big/ \partial \beta_2$ and $\beta_2= \frac{1}{k_BT_2}$. In the next stage of the cycle, the system remains in the bath at temperature $T_2$ and we remove the wall isothermally. The energy for this stage of the cycle is same as given in Eq. ~\eqref{g}. The  partition function  for the third stage of the cycle is 
\begin{equation}\label{eqnn7} \nonumber
Z_D = \sum_n e^{-\beta_2 E_n},
\end{equation}
where  $U_D$ can be similarly calculated as $U_C$. The heat exchanged for the third stage of the cycle (similar to Eq. ~\eqref{eqnn4}) stands as
\begin{equation}\label{eqnn8}
Q_{CD} \equiv U_D - U_C + k_B T_2 ln Z_D - k_B T_2 ln Z_C.
\end{equation}
Now, in the final stage of the cycle, the system reverts back to  the first stage of the cycle, i.e., the system is now connected to the heat bath at temperature $T_1$. The  energy exchange for the system when it reverts back to its initial stage is expressed as 
\begin{equation}\label{eqnn9}
Q_{DA} = U_A- U_D.
\end{equation}
We calculate the total work done for this cycle in terms of the uncertainty relation of the position and the momentum operator. It is evaluated using Eq.~\eqref{eqnn4}, ~\eqref{eqnn6}, ~\eqref{eqnn8} and ~\eqref{eqnn9} as
\begin{eqnarray}\label{eqnn10} \nonumber
W & \equiv & Q_{AB} + Q_{BC} + Q_{CD} + Q_{DA} \\ 
& = & \frac{8L^2\alpha}{\hbar^2 \pi^2} \Big[ f\, ln\Big(\frac{Z_B}{Z_A}\Big) + g \, ln\Big(\frac{Z_D}{Z_C}\Big) \Big]\, , 
\end{eqnarray}
where $f= \Big[\frac{16c\sqrt{2mc}}{\pi^3 \hbar^2}(\Delta X_{T_1} + \Delta P_{T_1} + C_{T_1})\Big]$ and $g = \Big[\frac{16c\sqrt{2mc}}{\pi^3 \hbar^2}(\Delta X_{T_2} + \Delta P_{T_2} + C_{T_2})\Big]$.
The efficiency of this engine from the thermal uncertainty relation standpoint using Eq.~\eqref{eqnn4}, ~\eqref{eqnn6}, ~\eqref{eqnn8} and ~\eqref{eqnn9} is
\begin{eqnarray} \label{eqnn11}\nonumber
\eta & \equiv & 1 + \frac{Q_{BC} + Q_{CD}}{Q_{DA} + Q_{AB}} \\ \nonumber
& = & \frac{ \Big(\bar{n}_{T_2}^2 \, ln\Big(\frac{Z_D}{Z_C}\Big) + \bar{n}_{T_1}^2 \, ln\Big(\frac{Z_B}{Z_A}\Big)\Big) }{\Big(-\bar{n}_{T_2}^2/2 + \bar{n}_{T_1}^2\Big(ln\Big(\frac{Z_B}{Z_A}\Big)+1/2\Big)\Big)} \\ 
& = & \frac{\Big[ g \, ln\Big(\frac{Z_D}{Z_C}\Big) + f \, ln\Big(\frac{Z_B}{Z_A}\Big) \Big]}{\Big[-g/2 +f ( ln\Big(\frac{Z_B}{Z_A}\Big) + 1/2) \Big]}\, . 
\end{eqnarray}
 
 \begin{figure}[t]
 \center
\includegraphics[width=1.0\columnwidth]{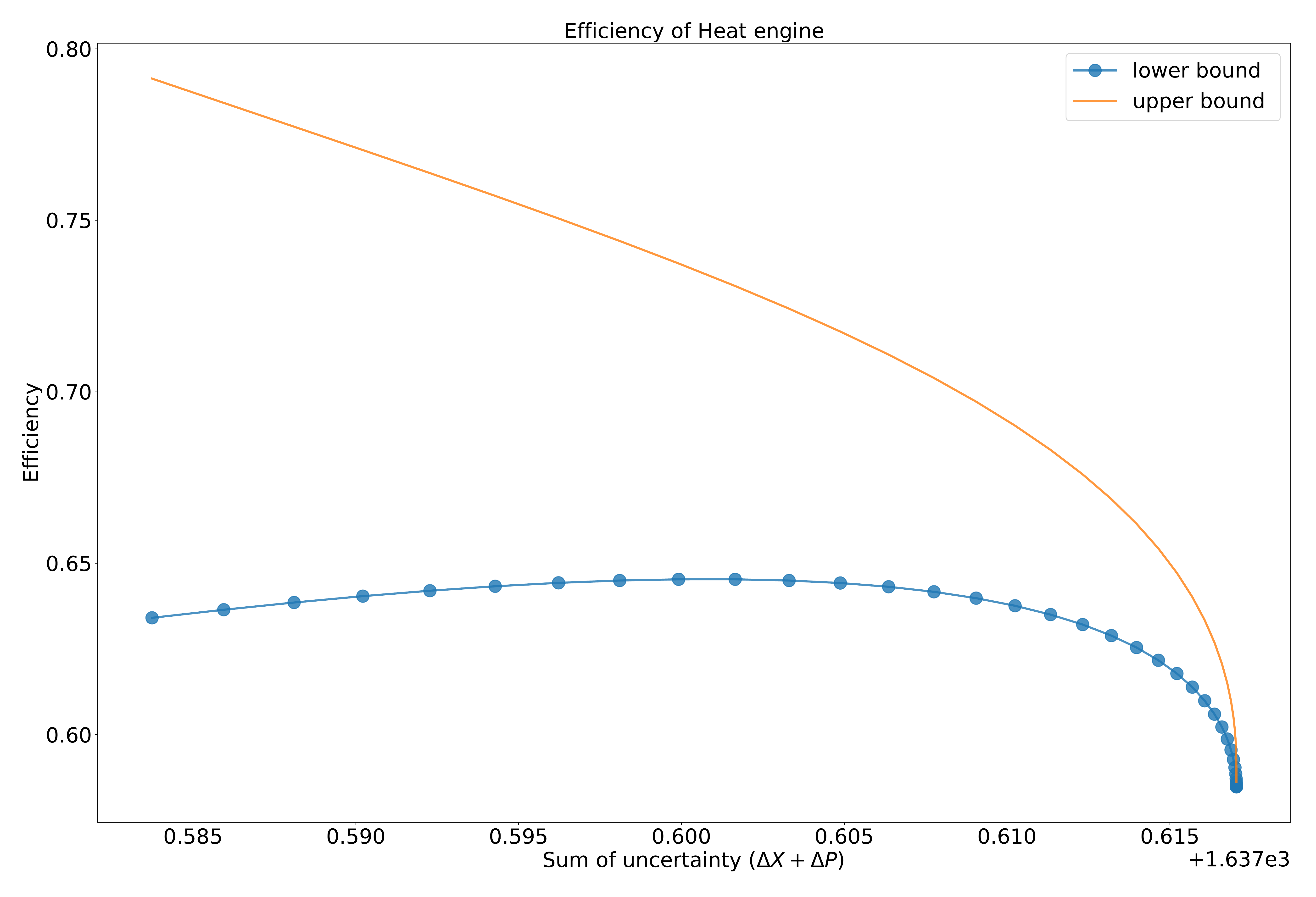}
\caption{The  efficiency bound for a relativistic model of heat engine.}
 \label{fig4}
\end{figure}

In Eq. \eqref{eqnn11}, we have evaluated the upper and the lower bound of the efficiency  with respect to the bound that we have analyzed  for the thermal uncertainty relation of the position and the momentum operator. Here, $f$ and $g$ provides the required uncertainty relation for the explanation of the bound of the efficiency. We can evaluate the lower bound of $f$ and $g$ in the  Eq. \eqref{eqnn11} from Eq. \eqref{eqn1} and its upper bound from Eq. \eqref{eqn4}.

Here, we have evaluated the relation between the efficiency of a heat engine for a relativistic particle with the variance of position  and momentum operator. The upper bound of the efficiency of the heat engine is monotonously decreasing function with the increase in temperature. From Fig. \ref{fig4}, we can infer that the variation of the lower bound with uncertainty is less for lower values of uncertainty, but there is a sudden dip when there is an increase in the uncertainty measure. The upper and lower bound  of the efficiency of the heat engine predicts the same rate of accuracy when the uncertainty takes higher value.

With the increase in the uncertainty, the conversion ratio of the heat engine decreases as the thermal energy of the system is directly proportional to the uncertainty of the system. In the case of the upper bound of the efficiency which is depicted in terms of the uncertainty relation defined in Eq. \ref{eqn8},  the decrease in the efficiency is more prominent due to the presence of the exponential component which causes exponential growth in the thermal energy of the engine and the dissipated heat over the work output. Whereas in the case of the lower bound we encounter a small variation of the efficiency for the lower value of the uncertainty. This can be easily analyzed from Eq. \ref{eqn2} where we encounter no exponential component which can depict a dominant effect on the thermal energy of the engine. If we equate the lower bound of the efficiency with the upper bound of the efficiency we encounter that it converges at high uncertainty. This show that for higher values of uncertainty the conversion ratio of the thermal energy to work reduces rapidly due to the steep growth in the thermal energy with higher uncertainty.

\section{Discussion and Conclusion}\label{section6}
Heat engine plays a key role for a better comprehension of quantum thermodynamics. In this work, we have considered a potential well model with a relativistic particle confined in it,  which acts as the working substance for the heat engine. Whether this can be globally extended to all the models that are considered for the analysis of heat engines and refrigerators is an open area to explore. 

We have given the analytic formulation of the work and efficiency of the engine in terms of the thermal uncertainty relation. Based on our formulation, the physical properties of the heat engine and the thermodynamic variables are as follows.

(a) The total work and the efficiency of the heat engine for the relativistic particle depends on the position and momentum of the particle. The variation in the uncertainty relation of the position and the momentum of the particle has a direct impact on the efficiency rate and the work of the engine. The upper bound of the efficiency of the engine drops gradually when the uncertainty of the observable increases, whereas the lower bound of the efficiency decreases when the variation in the uncertainty relation is high.

(b) Our formulation develops a direct connection of every quantum thermodynamic variable with the uncertainty relation. Helmholtz free energy for this relativistic system conveys the dependence of the internal energy of the system with the thermal uncertainty relation. The entropy which can be evaluated from Helmholtz free energy thus has a dependency on the uncertainty relation. The entropy of the system increases when the uncertainty of the incompatible observables increases for a definite temperature.

(c) The uncertainty relation is the cornerstone of quantum mechanics. Hereby applying this fundamental principle of quantum mechanics, we are able to predict the efficiency and the total work of the engine without performing any measurement. So, the measurement cost for the system gets reduced when we replace the classical model by a suitable quantum model, as has been done in this work.

All the well-known methods for the measurement of entanglement converges to the analysis of entropy~\cite{plen}. Now, if the system that is being analyzed can be modeled with a quantum model, we can study the entanglement property from the uncertainty relation viewpoint for the system.  If this method can explain the relativistic entanglement property, then this can act as a standard measure of entanglement. This might be a solution to the open problem of entanglement measure.  A parallel analysis of our defined model for the non relativistic regime is shown in our work \cite{pc12}.

This work can be further extended in the analysis of quantum engine in deformed space structures~\cite{kempf,ques,melj} through the relationship of generalized uncertainty relation (GUP) with the thermodynamic variables. In the paper~\cite{rezek}, the non-commutativity of the kinetic energy and the potential energy of quantum harmonic heat engine has been explored in great detail. So, analysis of heat engine in the deformed space structures~\cite{pc} is an open problem to explore in near future.

 Enhancement of entanglement in non-commutative space has been explored in details~\cite{deys,ghosh}. This raises a question whether deformed space structure can give a boost to the quantum engines under study. The holographic interpretation of entanglement entropy of anti-de Sitter (ADS)/conformal field theory (CFT) has been explored in the paper~\cite{ryu}. We can explore this from uncertainty viewpoint.
 
 One can also bridge a connection between the relativistic heat engines with the relativistic condensed matter physics. In some of the previous works~\cite{mani,choto,azimi,choto1,tul}, several approaches to design materials for non-relativistic engines and refrigerators are explored. Thus, it may also be possible to design materials for the analysis of the relativistic engines using the relativistic density functional theory~\cite{mac,ant,stra,sham1,guo,schr}. Cycles, when accompanied by the quantum phase transition, have a direct impact on the thermodynamic performance~\cite{may}. So, one possible application of our work could be to develop a connection between the uncertainty relations associated with the thermodynamics cycles with the quantum phase transition.

 The study of other thermodynamic cycles along with the analysis for developing the bound for different thermodynamic parameters is a wide open area to explore. Here, we have shown that entropy can be mapped with the uncertainty relation. This raises a question whether all thermodynamic variables and cycles can be mapped with the uncertainty of the observables for the working system under consideration.


\bibliographystyle{h-physrev4}


\end{document}